\documentclass[runningheads]{llncs}
\usepackage{graphicx}
\usepackage{multirow}
\usepackage{subcaption}
\usepackage{amsmath,amsfonts,amssymb}
\usepackage{MnSymbol}
\usepackage{algorithm}%
\usepackage[table,xcdraw]{xcolor}

\usepackage[indLines=false]{algpseudocodex}%
\algrenewcommand\algorithmicrequire{\textbf{Input:}}

 \mathchardef\mhyphen="2D

\begin{document}
\title{Imposing Rules in Process Discovery: an Inductive Mining Approach\thanks{\scriptsize This research was supported by the research training group ``Dataninja'' (Trustworthy AI for Seamless Problem Solving: Next Generation Intelligence Joins Robust Data Analysis) funded by the German federal state of North Rhine-Westphalia.}}
\titlerunning{Imposing Rules in Process Discovery}
\author{Ali Norouzifar \inst{1}\orcidID{0000-0002-1929-9992} \and
Marcus Dees\inst{2}\orcidID{0000-0002-6555-320X} \and
Wil van der Aalst\inst{1}\orcidID{0000-0002-0955-6940}}
\authorrunning{A. Norouzifar et al.}
\institute{RWTH University, Aachen, Germany \email{{ali.norouzifar, wvdaalst}@pads.rwth-aachen.de} \and
UWV Employee Insurance Agency, Amsterdam, Netherlands \\
\email{Marcus.Dees@uwv.nl}}
\maketitle              % typeset the header of the contribution
\setcounter{footnote}{0}
\begin{abstract}
 Process discovery aims to discover descriptive process models from event logs. These discovered process models depict the actual execution of a process and serve as a foundational element for conformance checking, performance analyses, and many other applications. While most of the current process discovery algorithms primarily rely on a single event log for model discovery, additional sources of information, such as process documentation and domain experts' knowledge, remain untapped. This valuable information is often overlooked in traditional process discovery approaches. In this paper, we propose a discovery technique incorporating such knowledge in a novel inductive mining approach. This method takes a set of user-defined or discovered rules as input and utilizes them to discover enhanced process models. Our proposed framework has been implemented and tested using several publicly available real-life event logs. Furthermore, to showcase the framework's effectiveness in a practical setting, we conducted a case study in collaboration with UWV, the Dutch employee insurance agency.

\end{abstract}

\section{Introduction}
Process discovery seeks to identify process models that provide the most accurate representation of a given process. The quality of discovered process models is measured using evaluation metrics while ensuring comprehensibility for human understanding and alignment with domain experts' knowledge. Many state-of-the-art discovery approaches rely solely on event logs as their primary source of information, often neglecting additional valuable resources such as the knowledge of process experts and documentation detailing the process \cite{DBLP:journals/cii/Beerepoot}. These overlooked resources can significantly enhance the discovery process \cite{DBLP:journals/cii/SchusterZA22}.

Process experts often possess a common understanding of how the process functions and additional resources like process diagrams may be available alongside event logs. This information can often be expressed in human language, e.g., activities $a$ and $b$ cannot occur together, or activity $a$ cannot occur after activity $b$. Automated methods can also be used to discover such relations between activities from an event log \cite{DBLP:conf/caise/MaggiBA12,DBLP:journals/tmis/CiccioM15}. We propose a novel framework to impose such information in process discovery. Our framework leverages Inductive Mining (IM) techniques, with a distinctive feature that allows it to incorporate a set of rules as an additional input. We assume the rules are given by an oracle, e.g., user-defined or discovered rules using automated methods. The framework is designed independently from the source that provides the rules.

\begin{figure}[htb]
\begin{subfigure}{0.54\textwidth}
\centering
\includegraphics[scale=0.35]{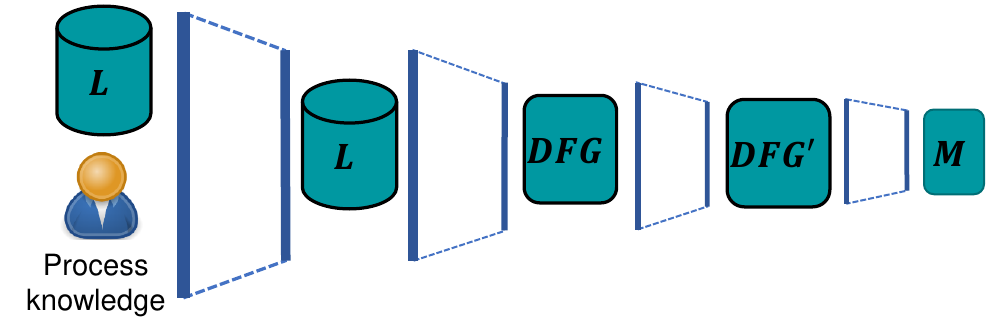}
\caption{\small Information flow in traditional IM frameworks.}
\label{information_loss1}
\end{subfigure} \;\;
\begin{subfigure}{0.45\textwidth}
\centering
\includegraphics[scale=0.35]{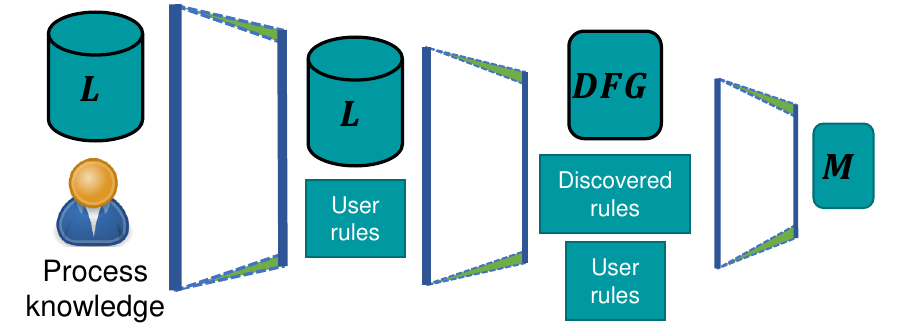}
\caption{\small Information flow in our proposed framework.}
\label{information_loss2}
\end{subfigure}
\caption{\small Comparing the information flow in IM frameworks with our framework.}
\label{information_loss}
\vspace{-10pt}
\end{figure}

IM techniques discover block-structured process models that are both comprehensible for humans and offer guarantees, such as soundness~\cite{DBLP:conf/bpm/LeemansFA13}. The information flow in one recursion of IM approaches is illustrated in Fig.~\ref{information_loss1}. Overlooking other information sources, shown as process knowledge, results in some information loss. In each recursion of the IM techniques, a Directly Follows Graph (DFG) is derived from the event log that serves as the basis for determining the process structure. The conversion of an event log to a DFG can lead to information loss. Some variants include filtering mechanisms to eliminate infrequent behaviors ($DFG^{\prime}$) which contribute to more information loss.  To avoid potential blocks in IM techniques when no cut is detected, heuristics are used to prevent the discovery process from becoming impeded, albeit at the cost of potentially generating over-generalized models. 
 
\vspace{-10pt}

In this paper, we present an inductive mining framework designed to leverage encoded process knowledge expressed in the \textit{Declare} language. This additional information is utilized to enhance the discovery of improved process models, as depicted in Fig.~\ref{information_loss2}. The information we can preserve in each step is illustrated in green color. The process knowledge can be encoded as user-defined rules, providing an extra source of information alongside the event log. Automated declarative process discovery methods can be employed to reveal process structures, compensating for the information loss in the extracted DFG. Importantly, our approach does not rely on fall-throughs. In cases where no perfect cut exists, it returns the most promising one based on specific cost functions similar to~\cite{DBLP:conf/sac/NorouzifarA23}.

\section{Related Work}
The incorporation of process knowledge or additional information sources in process discovery has been explored in various formats~\cite{DBLP:journals/cii/SchusterZA22}. This information may originate from domain experts' knowledge, business documents, or be derived through automated algorithms. Utilizing such information before the actual discovery process to preprocess event logs aligns with common steps in data analysis frameworks. However, the involvement of these information sources in process discovery remains limited within the literature.

In \cite{DBLP:journals/jmlr/GoedertierMVB09}, the authors proposed an automatic approach for discovering artificially created events, revealing aspects not observed in the event log. This information guides the discovery algorithm towards generating more robust process models. Another approach, presented in \cite{DBLP:conf/icsoc/RembertOMG13}, involves using prior knowledge to learn a control flow model in the form of information control nets. In \cite{Yahya2013} a method is introduced that leverages user knowledge in the form of relations between activities to construct a directly follows graph. Unlike \cite{DBLP:conf/icsoc/RembertOMG13} and \cite{Yahya2013} our focus is on Petri net models. Another strategy involves post-discovery process model repair based on predefined preferences \cite{DBLP:conf/bpm/FahlandA12}. Additionally, interactive process discovery and online process discovery are explored as related works \cite{DBLP:journals/softx/SchusterZA23}.

Declare language \cite{DBLP:conf/caise/MaggiBA12,DBLP:journals/tmis/CiccioM15} and compliance rule graph language \cite{DBLP:conf/er/KnupleschRLKR13} exemplify rule modeling techniques that offer high interpretability. In \cite{DBLP:conf/simpda/DixitBAHB15}, the use of declarative rules provided by users or discovered by declarative mining algorithms is considered to enhance the quality of discovered process models. The rules are not used directly in the discovery, instead a discovered model is used, several modifications are applied and the best model that adheres to the rules is selected. In our proposed method, we directly use the rules in process discovery. 

Automatic process discovery is a crucial research area, yet fundamental questions persist despite the plenty of proposed algorithms. Two types of inductive mining algorithms exist in the literature: those that output a unique cut in each recursion without quality evaluation \cite{DBLP:conf/bpm/LeemansFA13}, and those that select the best cut over a set of candidates based on quality measures \cite{DBLP:conf/sac/NorouzifarA23}, \cite{DBLP:conf/icpm/DettenSL23}, and \cite{DBLP:conf/icpm/BronsSF21}.  We extend the idea proposed in  \cite{DBLP:conf/sac/NorouzifarA23} and make it capable of using rules in discovery recursions.  

\vspace{-5pt}
\section{Motivating Examples}
\begin{figure}[htb]
\begin{subfigure}{0.6\textwidth}
\centering
\includegraphics[scale=0.15]{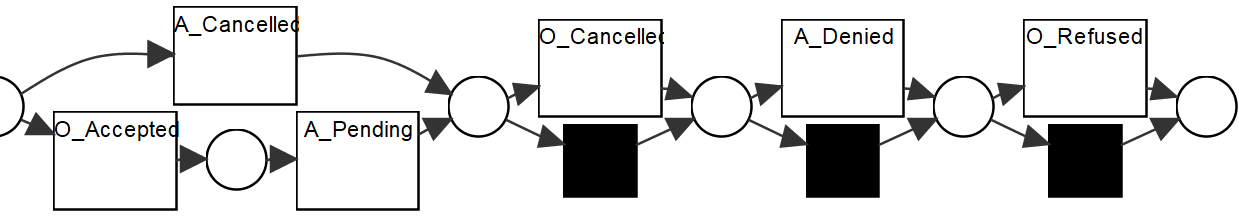}
\caption{\small BPIC 2017, discovered model using IMf with $f{=}0.2$.}
\label{motiv_example_before_BPIC17}
\end{subfigure}\;\;
\begin{subfigure}{0.35\textwidth}
\centering
\includegraphics[scale=0.15]{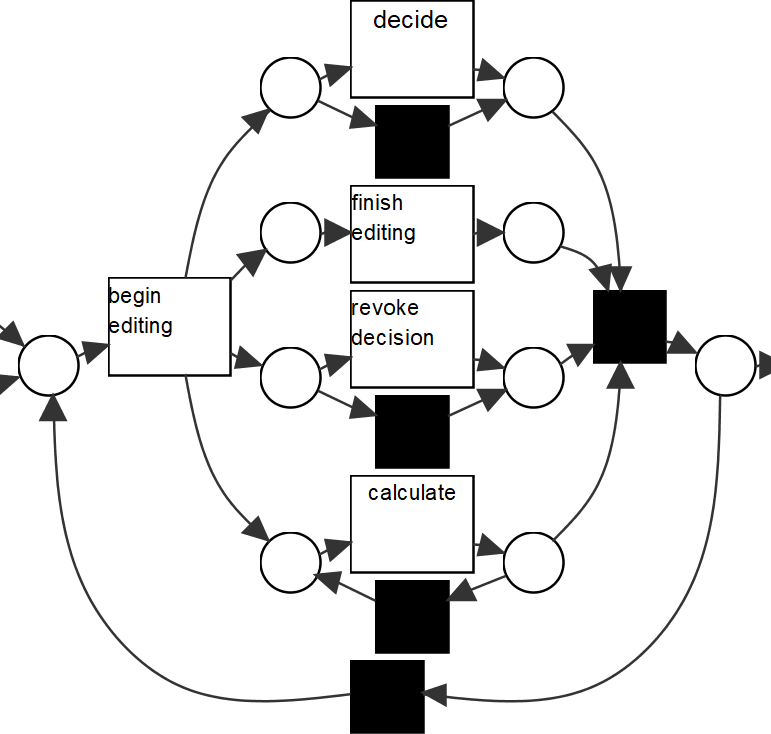}
\caption{\small BPIC 2018, discovered model using IMf with $f{=}0.2$.}
\label{motiv_example_before_BPIC18}
\end{subfigure}\\
\begin{subfigure}{1\textwidth}
\centering
\includegraphics[scale=0.13]{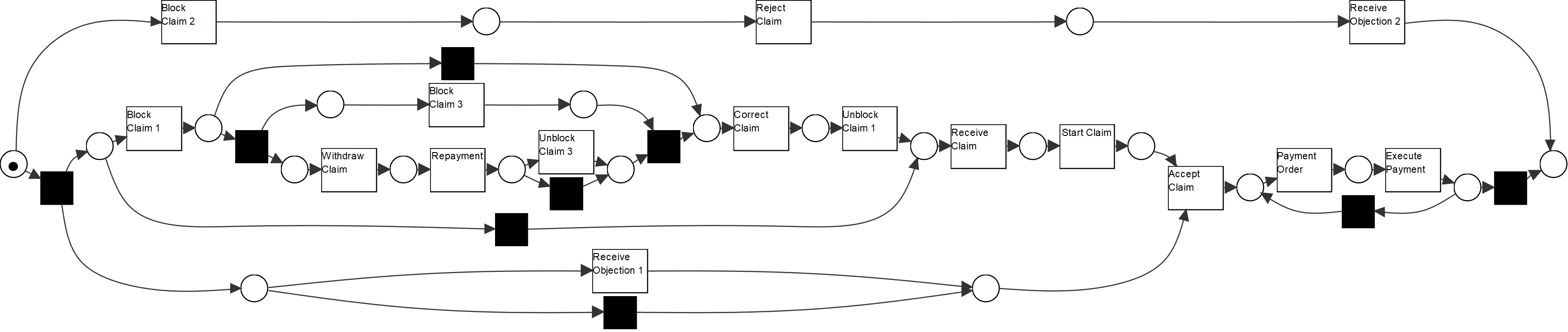}
\caption{\small UWV, discovered model using IMf with $f=0.2$.}
\label{IMf_model_uwv}
\end{subfigure}
\caption{Motivating examples, using IMf to discover process models for BPIC 2017, BPIC 2018, UWV event log.}
\label{motiv_example_before}
\end{figure}

\vspace{-5pt}
To motivate the research question addressed in this paper, we offer examples highlighting the necessity of our investigation. Figures~\ref{motiv_example_before_BPIC17} and~\ref{motiv_example_before_BPIC18} showcase a part of Petri net models discovered from publicly available real-life event logs, i.e., BPIC 2017 and BPIC 2018, using IMf algorithm with 0.2 as the infrequency filtering parameter. Additionally, Figure~\ref{IMf_model_uwv} exemplifies a process model discovered using the same settings from an event log provided by UWV agency consisting of cases related to a claim handling process which is investigated in detail in the evaluation section.

In Fig.~\ref{motiv_example_before_BPIC17}, despite the sequential relation between final states identified by IMf, the event log analysis reveals that only one of the final states can occur which means either an application is accepted (\textit{A\_Submitted}), canceled (\textit{A\_Cancelled}), or denied by the client (\textit{A\_Denied}). In Fig.~\ref{motiv_example_before_BPIC18}, the model overgeneralizes the observed behavior and allows for some ordering of the activities which does not make sense both based on the event log analysis and a common sense we have based on the activity names. After \textit{begin editing}, \textit{calculate} should occur, followed by \textit{finish editing} which makes a case ready to make a decision (activity \textit{decide}). Activity \textit{revoke decision} may then occur after making a decision. A case can have multiple repetitions of the explained procedure. The shown model allows for many behaviors that deviate from this procedure, e.g., \textit{decide} before \textit{calculate} and \textit{finish editing}, or \textit{revoke decision} before \textit{decide}.

In the UWV event log, based on the domain knowledge a case must start with \textit{Receive Claim} and \textit{Start Claim}, however, in Fig.~\ref{IMf_model_uwv}, the process model allows for many activities before receiving a claim that does not make sense. \textit{Block Claim 1}, \textit{Block Claim 2}, and \textit{Block Claim 3} have specific meanings in this process. The ordering of these activities does not make sense according to our investigations with process experts at UWV, e.g., \textit{Block Claim 1} can only occur after starting a claim when some information is missing and should be followed by \textit{Correct Claim} and \textit{Unblock Claim 1} to make it ready to get accepted (\textit{Accept Claim}). Figures \ref{motive_IMr_2017}, \ref{motive_IMr_2018}, and \ref{uwv_model_after} show the models discovered by our proposed framework for BPIC 2017, BPIC 2018, and UWV event logs respectively.

\vspace{-10pt}

\vspace{-30pt}
\section{Preliminaries}
Considering ${\mathcal{A}}$ as the universe of activities, $s {\in} \mathcal{A}^{*}$ denotes a sequence of activities where $s(i)$ indicates the $i$-th element in this sequence. $\mathcal{P}(\Sigma)$ denotes the power set over set $\Sigma \subseteq \mathcal{A}$. 
% Process discovery starts from event data which is extracted from information systems. 
We introduce an event log formally as a multiset of traces, i.e., a sequence of activities.

\begin{definition}[Event log]
Let  ${\mathcal{A}}$ be the universe of activities. A trace  $\sigma = \langle a_1, a_2, \ldots, a_n \rangle  \in {\mathcal{A}}^*$ is a finite sequence of activities. Each occurrence of an activity in a trace is an event. An event log $L\in \mathcal{B}({\mathcal{A}}^*)$ is a multiset of traces.  $\mathcal{L}$ is the universe of event logs. 
\end{definition}

Since we are building our framework based on the inductive mining technique, process tree notation is used as the representation. Process trees can be converted to Petri nets or BPMN models as more popular process notations. 

\begin{definition}[Process tree]
Let $\bigoplus=\lbrace \rightarrow,\times, \wedge, \circlearrowleft \rbrace$ be the set of process tree operators and let $\tau \not \in {\mathcal{A}}$ be the so-called silent transition, then
\begin{itemize}
\item activity $a\in {\mathcal{A}}$ is a process tree,
\item the silent activity $\tau$ is a process tree,
\item  let $M_1, \cdots, M_n$ with $n>0$ be process trees and let $\oplus \in \bigoplus$ be a process tree operator, then
$\oplus(M_1, \cdots, M_n)$ is a process tree.
\end{itemize}
$\mathcal{M}_{\Sigma}$ is the set of all possible process trees generated over a set of activities $\Sigma {\subseteq} \mathcal{A}$.
\end{definition}

Each process tree operator $\oplus \in \lbrace \rightarrow,\times, \wedge, \circlearrowleft \rbrace$ has a semantic which generates a special type of behavior. 
The function $\phi: \mathcal{M}_{\Sigma} \rightarrow \mathcal{P}(\Sigma^*)$ extracts the set of traces allowed by a process tree which we refer to as the language of this process tree. If a process tree consists of an operator as the root node and single activities as children, the language of it is as follows:
\begin{itemize}
    \item $\rightarrow$ denotes the sequential composition of children, e.g., {\small$\phi(\rightarrow(a,b)) {=} \{\langle a,b \rangle \}$}.
    \item $\times$ represents the exclusive choice between children, e.g., {\small$\phi(\times(a,b)) {=} \{\langle a \rangle, \langle b \rangle \}$}
    \item $\wedge$ denotes the concurrent composition of children, e.g., {\small$\phi(\wedge(a,b)) {=} \{\langle a,b \rangle, \langle b,a \rangle \}$}
    \item $\circlearrowleft$ represents the loop execution in which the first child is the body of the loop and the other children are redo children, e.g., {\small $\phi(\circlearrowleft(a,b)) {=} \{\langle a \rangle, \langle a,b,a \rangle, ... \}$}.
\end{itemize}

$M = \times(\rightarrow(a,b),\wedge(\times(c,\tau),d))$ is a more complex example such that $\phi(M)= \{\langle a,b \rangle,$ $\langle d \rangle,\langle c,d \rangle,\langle d,c \rangle \}$.

Consider $\mathcal{G}(L)$ as a function that extracts the directly follows graph $(\Sigma,E)$ from event log $L \in \mathcal{L}$ such that $\Sigma = \{a \in \sigma \vert \sigma \in L\}$ is the set of activities and
$E=\{(a_1,a_2) \vert \exists \sigma {\in} L \wedge 1 {\leq} i{<}\vert \sigma \vert: \sigma(i) {=} a_1 \wedge  \sigma(i+1) {=} a_2 \}$ is the set of edges.

\section{Inductive Miner with Rules (IMr)}

\begin{figure}[htb]
    \centering
        \begin{minipage}{0.47\textwidth}
        \centering
        \includegraphics[width=0.9\linewidth]{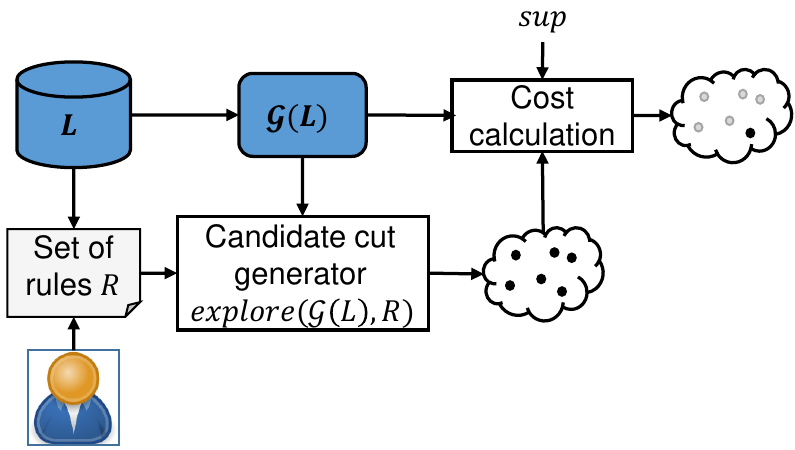}
        \caption{\small One iteration of IMr, the framework proposed in this paper, identifies candidate cuts adhering to specified rules and selects the cut with minimum cost, incorporating cost functions from \cite{DBLP:conf/sac/NorouzifarA23}.}
        \label{discovery_overview}
    \end{minipage}\;
    \begin{minipage}{0.51\textwidth}
        \begin{algorithm}[H]
            \scriptsize
            % \footnotesize
            \caption{\small $IMr$ algorithm}
            \label{IM-rules}
            \begin{algorithmic}
            \Function{$IMr$}{$L,sup,R$}
                 \LComment{$L {\in} {\mathcal{L}}$ is an event log, $sup \in [0,1]$ is a process discovery parameter and $R$ is the set of rules.}
                \State $base = checkBaseCase(L,sup)$
                \If {$checkBaseCase$ successful}
                    \State \Return $base$
                \EndIf
                \State $C=explore(\mathcal{G}(L),R)$
                \State $(\oplus,\Sigma_1,\Sigma_2)=\underset{c \in C}{\arg\min} \lbrace ov\_cost_{\mathcal{G}(L)}(c, sup) \rbrace$
                \State $L_1, L_2 = SPLIT(L,(\oplus,\Sigma_1,\Sigma_2))$
                \State \Return $\oplus(IMr(L_1,sup,R),IMr(L_2, sup,R))$
                \EndFunction
            \end{algorithmic}
        \end{algorithm}
    \end{minipage}
\end{figure}

In this paper, we adapt and extend the IMbi framework proposed in \cite{DBLP:conf/sac/NorouzifarA23} to allow for rules being used in process discovery. The new framework is referred to as IMr in this paper. In Fig.~\ref{discovery_overview}, the main idea of this paper is illustrated. The IMbi framework, designed for two event logs, a desirable event log and an undesirable event log, is adapted in our approach. For the sake of simplicity, the undesirable event log is excluded\footnote{The parameter $ratio$ controls the relevance of the undesirable event log; setting the parameter $ratio=0$ disregards $L^-$, focusing solely on the desirable event log $L^+$.}, however, the approach is adaptable to scenarios where the undesirable event log is included. The algorithm finds binary cuts in each recursion like IMbi.

\begin{definition}[Binary Cut]
Let  $L {\in} {\mathcal{L}}$ be an event log. $\mathcal{G}(L){=}(\Sigma, E)$ is the corresponding DFG. A binary cut $(\oplus,\Sigma_1,\Sigma_2)$ divides $\Sigma$ into two partitions, such that $\Sigma_1 {\cup} \Sigma_2 {=} \Sigma$, $\Sigma_1  {\cap} \Sigma_2 {=} \emptyset$, and $\oplus {\in} \lbrace \rightarrow,\times, \wedge, \circlearrowleft \rbrace$ is a cut type operator.
\end{definition}

Algorithm~\ref{IM-rules} shows how IMr works. In each recursion, $explore(\mathcal{G}(L), R)$ explores the DFG extracted from event log $L$ and returns a set of candidate cuts. We explain in this paper how the set of rules $R$ can be used to prune the set of candidate cuts. We use the $ov\_cost$ function as defined in \cite{DBLP:conf/sac/NorouzifarA23} to compare the cuts. 
The cost value for each candidate cut is determined by counting the number of deviating edges and estimating the number of missing edges required to modify $\mathcal{G}(L)$ to align with the candidate cut.
Parameter $sup \in [0,1]$ specifies to what extent missing behaviors should be penalized. Among the set of candidate cuts, the cut with the minimum cost is selected.  The algorithm continues with splitting the event log based on the selected cut (function $SPLIT$) and proceeding to the next recursion.

\vspace{-20pt}
\vspace{-20pt}

\vspace{-10pt}
\subsection{The Set of Rules}
The main difference between IMbi and IMr is the use of the set of rules $R$ in finding the set of candidate cuts. Each rule $r {\in} R$ is a constraint that limits the behavior.
Process models may allow for traces that violate or satisfy a rule. We need a formal language to implement the idea, therefore, without loss of generality, we use declarative constraints in this paper.  However, the concept is general, and any rule with clear semantics can be employed, provided that \textit{there exists a clear mapping between rule satisfaction and the allowance of certain cut types}.

% \begin{definition}[Rules]
% Let $\mathcal{U}_A$ be the universe of activities. A rule $r \in \mathcal{U}_A \times template$ uses a template and 
% \end{definition}

A declarative constraint is an instantiation of a template that involves one or more activities~\cite{DBLP:conf/caise/MaggiBA12,DBLP:journals/tmis/CiccioM15}. Templates are abstract parameterized patterns. In this paper, a subset of declarative templates is used including:
\begin{itemize}
    \item $at \mhyphen most(a)$: $a$ occurs at most once.
    \item $existence(a)$: $a$ occurs at least once.
    \item $response(a,b)$: If $a$ occurs, then $b$ occurs after $a$.
    \item $precedence(a,b)$: $b$ occurs only if preceded by $a$.
    \item $co\mhyphen existence(a,b)$: $a$ and $b$ occur together.
    \item $not\mhyphen co\mhyphen existence(a,b)$: $a$ and $b$ never occur together.
    \item $not\mhyphen succession(a,b)$: $b$ cannot occur after $a$. 
    \item $responded\mhyphen existence(a,b)$: If $a$ occurs in the trace, then $b$ occurs as well.
\end{itemize}
Including other declarative templates requires more sophisticated design choices and considerations, therefore, we only focus on a subset of them. Process experts can encode their knowledge and understanding in the form of declarative rules and use them in process discovery as we explain in this paper. In addition to user-defined rules, automated declarative process discovery algorithms such as Declare Miner \cite{DBLP:conf/caise/MaggiBA12} and MINERful \cite{DBLP:journals/tmis/CiccioM15} can be used to discover declarative constraints from event logs. If trace $\sigma$ violates constraint $r$, we show it as $\sigma \nvDash r$. For example, consider $\sigma = \langle a, b, a,c \rangle$, and $r = response(a,b)$.  $\sigma \nvDash r$ because the second $a$ in trace $\sigma$ is not followed by a $b$.

\begin{definition}[Constraint violation]
    Let $L {\in} \mathcal{L}$ be an event log and $\mathcal{G}(L)=(\Sigma,E)$ be the extracted DFG, and $R$ be the set of rules. $c \nvDash r$ denotes that cut $c = (\oplus,\Sigma_1, \Sigma_2) \in explore(\mathcal{G}(L),R)$ violates the constraint $r\in R$, meaning that for all process trees $M{\in} \mathcal{M}_c$ where $\mathcal{M}_c=\{ \oplus(M_1,M_2) \vert M_1 \in \mathcal{M}_{\Sigma_1} \wedge M_2 \in \mathcal{M}_{\Sigma_2} \}$, there is a trace $\sigma {\in} \phi(M)$ such that $\sigma \nvDash r$. 
\end{definition}

For example, $c{=}(\rightarrow,\{b\}, \{a\})$ violates rule $response(a,b)$, since all the models $M \in \mathcal{M}_c$ allow for trace $\langle b,a \rangle$ which violates $response(a,b)$.

\subsection{Candidate Cuts Pruning}
Consider $R$ as the set of all rules that are given by the user or are discovered using declarative process discovery algorithms. We remove a cut $c$ from the set of candidate cuts $explore(\mathcal{G}(L),R)$ if there is a rule $r\in R$ such that $c \nvDash r$. In Table~\ref{one_rules_allowance}, it is shown with red color for each single activity constraint which rules should be rejected. Similarly, in Table~\ref{two_rules_allowance}, for each two activities constraint, it is shown which rules should be rejected. Next, we explain in more detail how the cut-pruning algorithm works. Please note that although declarative rules may capture certain long-term dependencies, our discovery algorithm's representational bias might fail to adequately represent them. Consequently, the pruned candidate set may not contain any rules. In such cases, the algorithm identifies the set of candidate cuts without considering the provided rule set. Further elaboration on this phenomenon can be found in Section~\ref{open_chal}.

\vspace{-10pt}
\subsubsection*{$at \mhyphen most(a)$} $a$ occurs at most once.
\begin{itemize}
    \item {\small $c{=}(\circlearrowleft,\Sigma_1,\Sigma_2) {\nvDash} at \mhyphen most(a)$} if $a {\in} \Sigma_1$ ($a {\in} \Sigma_2$), because for any process tree $M{\in}\mathcal{M}_{c}$, there is a trace $\sigma {\in} \phi(M)$ which has multiple occurrences of activity $a$ because the body (the redo part) of loop process trees can occur several times.
    \vspace{-10pt}
\end{itemize}

\begin{table}[htb]
\caption{\small The cuts that should be rejected are shown with red color for constraints with one activity.}
\label{one_rules_allowance}
\centering
\begin{tabular}{l|ll|ll|ll|ll|}
\cline{2-9}
                                   & \multicolumn{2}{c|}{$(\rightarrow,\Sigma_1,\Sigma_2)$}                    & \multicolumn{2}{c|}{$(\times,\Sigma_1,\Sigma_2)$}                         & \multicolumn{2}{c|}{$(\wedge,\Sigma_1,\Sigma_2)$}                         & \multicolumn{2}{c|}{$(\circlearrowleft,\Sigma_1,\Sigma_2)$}                                      \\ \cline{2-9} 
                                   & \multicolumn{1}{l|}{$a \in \Sigma_1$}         & $a \in \Sigma_2$         & \multicolumn{1}{l|}{$a \in \Sigma_1$}         & $a \in \Sigma_2$         & \multicolumn{1}{l|}{$a \in \Sigma_1$}         & $a \in \Sigma_2$         & \multicolumn{1}{l|}{$a \in \Sigma_1$}         & $a \in \Sigma_2$                                \\ \hline
\multicolumn{1}{|l|}{at-most(a)}  & \multicolumn{1}{l|}{\cellcolor[HTML]{FFFFFF}} & \cellcolor[HTML]{FFFFFF} & \multicolumn{1}{l|}{\cellcolor[HTML]{FFFFFF}} & \cellcolor[HTML]{FFFFFF} & \multicolumn{1}{l|}{\cellcolor[HTML]{FFFFFF}} & \cellcolor[HTML]{FFFFFF} & \multicolumn{1}{l|}{\cellcolor[HTML]{FE0000}} & \cellcolor[HTML]{FE0000}{\color[HTML]{FE0000} } \\ \hline
\multicolumn{1}{|l|}{existence(a)} & \multicolumn{1}{l|}{\cellcolor[HTML]{FFFFFF}} & \cellcolor[HTML]{FFFFFF} & \multicolumn{1}{l|}{\cellcolor[HTML]{FE0000}} & \cellcolor[HTML]{FE0000} & \multicolumn{1}{l|}{\cellcolor[HTML]{FFFFFF}} & \cellcolor[HTML]{FFFFFF} & \multicolumn{1}{l|}{\cellcolor[HTML]{FFFFFF}} & \cellcolor[HTML]{FE0000}{\color[HTML]{FE0000} } \\ \hline
\end{tabular}
\end{table}
\vspace{-10pt}

\begin{table}[htb]
\caption{\small The cuts that should be rejected are shown with red color for constraints with two activities.}
\label{two_rules_allowance}
\centering
\begin{tabular}{l|llll|llll|}
\cline{2-9}
                                               & \multicolumn{4}{c|}{$(\rightarrow,\Sigma_1,\Sigma_2)$}                                                                                                                                                                                                                                                                                                                                & \multicolumn{4}{c|}{$(\times,\Sigma_1,\Sigma_2)$}                                                                                                                                                                                                                                                                                                                                      \\ \cline{2-9} 
                                               & \multicolumn{1}{l|}{\begin{tabular}[c]{@{}l@{}}$a \in \Sigma_1$\\ $b \in \Sigma_1$\end{tabular}} & \multicolumn{1}{l|}{\begin{tabular}[c]{@{}l@{}}$a \in \Sigma_2$\\ $b \in \Sigma_2$\end{tabular}} & \multicolumn{1}{l|}{\begin{tabular}[c]{@{}l@{}}$a \in \Sigma_1$\\ $b \in \Sigma_2$\end{tabular}} & \begin{tabular}[c]{@{}l@{}}$a \in \Sigma_2$\\ $b \in \Sigma_1$\end{tabular} & \multicolumn{1}{l|}{\begin{tabular}[c]{@{}l@{}}$a \in \Sigma_1$\\ $b \in \Sigma_1$\end{tabular}} & \multicolumn{1}{l|}{\begin{tabular}[c]{@{}l@{}}$a \in \Sigma_2$\\ $ b \in \Sigma_2$\end{tabular}} & \multicolumn{1}{l|}{\begin{tabular}[c]{@{}l@{}}$a \in \Sigma_1$\\ $b \in \Sigma_2$\end{tabular}} & \begin{tabular}[c]{@{}l@{}}$a \in \Sigma_2$\\ $b \in \Sigma_1$\end{tabular} \\ \hline
\multicolumn{1}{|l|}{response(a,b)}            & \multicolumn{1}{l|}{\cellcolor[HTML]{FFFFFF}}                                                    & \multicolumn{1}{l|}{}                                                                            & \multicolumn{1}{l|}{}                                                                            & \cellcolor[HTML]{FE0000}{\color[HTML]{FE0000} }                             & \multicolumn{1}{l|}{}                                                                            & \multicolumn{1}{l|}{}                                                                             & \multicolumn{1}{l|}{\cellcolor[HTML]{FE0000}}                                                    & \cellcolor[HTML]{FE0000}                                                    \\ \hline
\multicolumn{1}{|l|}{precedence(a,b)}          & \multicolumn{1}{l|}{}                                                                            & \multicolumn{1}{l|}{}                                                                            & \multicolumn{1}{l|}{}                                                                            & \cellcolor[HTML]{FE0000}{\color[HTML]{FE0000} }                             & \multicolumn{1}{l|}{}                                                                            & \multicolumn{1}{l|}{}                                                                             & \multicolumn{1}{l|}{\cellcolor[HTML]{FE0000}}                                                    & \cellcolor[HTML]{FE0000}                                                    \\ \hline
\multicolumn{1}{|l|}{co-existence(a,b)}        & \multicolumn{1}{l|}{}                                                                            & \multicolumn{1}{l|}{}                                                                            & \multicolumn{1}{l|}{}                                                                            &                                                                             & \multicolumn{1}{l|}{}                                                                            & \multicolumn{1}{l|}{}                                                                             & \multicolumn{1}{l|}{\cellcolor[HTML]{FE0000}}                                                    & \cellcolor[HTML]{FE0000}                                                    \\ \hline
\multicolumn{1}{|l|}{not-co-existence(a,b)}    & \multicolumn{1}{l|}{}                                                                            & \multicolumn{1}{l|}{}                                                                            & \multicolumn{1}{l|}{\cellcolor[HTML]{FE0000}}                                                    & \cellcolor[HTML]{FE0000}{\color[HTML]{FE0000} }                             & \multicolumn{1}{l|}{}                                                                            & \multicolumn{1}{l|}{}                                                                             & \multicolumn{1}{l|}{}                                                                            &                                                                             \\ \hline
\multicolumn{1}{|l|}{not-succession(a,b)}      & \multicolumn{1}{l|}{}                                                                            & \multicolumn{1}{l|}{}                                                                            & \multicolumn{1}{l|}{\cellcolor[HTML]{FE0000}}                                                    &                                                                             & \multicolumn{1}{l|}{}                                                                            & \multicolumn{1}{l|}{}                                                                             & \multicolumn{1}{l|}{}                                                                            &                                                                             \\ \hline
\multicolumn{1}{|l|}{responded-existence(a,b)} & \multicolumn{1}{l|}{}                                                                            & \multicolumn{1}{l|}{}                                                                            & \multicolumn{1}{l|}{}                                                                            &                                                                             & \multicolumn{1}{l|}{}                                                                            & \multicolumn{1}{l|}{}                                                                             & \multicolumn{1}{l|}{\cellcolor[HTML]{FE0000}}                                                    & \cellcolor[HTML]{FE0000}                                                    \\ \hline
\end{tabular}

\begin{tabular}{l|llll|llll|}
\cline{2-9}
                                               & \multicolumn{4}{c|}{$(\wedge,\Sigma_1,\Sigma_2)$}                                                                                                                                                                                                                                                                                                                                     & \multicolumn{4}{c|}{$(\circlearrowleft,\Sigma_1,\Sigma_2)$}                                                                                                                                                                                                                                                                                                                            \\ \cline{2-9} 
                                               & \multicolumn{1}{l|}{\begin{tabular}[c]{@{}l@{}}$a \in \Sigma_1$\\ $b \in \Sigma_1$\end{tabular}} & \multicolumn{1}{l|}{\begin{tabular}[c]{@{}l@{}}$a \in \Sigma_2$\\ $b \in \Sigma_2$\end{tabular}} & \multicolumn{1}{l|}{\begin{tabular}[c]{@{}l@{}}$a \in \Sigma_1$\\ $b \in \Sigma_2$\end{tabular}} & \begin{tabular}[c]{@{}l@{}}$a \in \Sigma_2$\\ $b \in \Sigma_1$\end{tabular} & \multicolumn{1}{l|}{\begin{tabular}[c]{@{}l@{}}$a \in \Sigma_1$\\ $b \in \Sigma_1$\end{tabular}} & \multicolumn{1}{l|}{\begin{tabular}[c]{@{}l@{}}$a \in \Sigma_2$\\ $ b \in \Sigma_2$\end{tabular}} & \multicolumn{1}{l|}{\begin{tabular}[c]{@{}l@{}}$a \in \Sigma_1$\\ $b \in \Sigma_2$\end{tabular}} & \begin{tabular}[c]{@{}l@{}}$a \in \Sigma_2$\\ $b \in \Sigma_1$\end{tabular} \\ \hline
\multicolumn{1}{|l|}{response(a,b)}            & \multicolumn{1}{l|}{\cellcolor[HTML]{FFFFFF}}                                                    & \multicolumn{1}{l|}{\cellcolor[HTML]{FFFFFF}}                                                    & \multicolumn{1}{l|}{\cellcolor[HTML]{FE0000}}                                                    & \cellcolor[HTML]{FE0000}{\color[HTML]{FE0000} }                             & \multicolumn{1}{l|}{\cellcolor[HTML]{FFFFFF}}                                                    & \multicolumn{1}{l|}{\cellcolor[HTML]{FFFFFF}}                                                     & \multicolumn{1}{l|}{\cellcolor[HTML]{FE0000}}                                                    & \cellcolor[HTML]{FFFFFF}                                                    \\ \hline
\multicolumn{1}{|l|}{precedence(a,b)}          & \multicolumn{1}{l|}{}                                                                            & \multicolumn{1}{l|}{\cellcolor[HTML]{FFFFFF}}                                                    & \multicolumn{1}{l|}{\cellcolor[HTML]{FE0000}}                                                    & \cellcolor[HTML]{FE0000}{\color[HTML]{FE0000} }                             & \multicolumn{1}{l|}{\cellcolor[HTML]{FFFFFF}}                                                    & \multicolumn{1}{l|}{\cellcolor[HTML]{FFFFFF}}                                                     & \multicolumn{1}{l|}{\cellcolor[HTML]{FFFFFF}}                                                    & \cellcolor[HTML]{FE0000}                                                    \\ \hline
\multicolumn{1}{|l|}{co-existence(a,b)}        & \multicolumn{1}{l|}{}                                                                            & \multicolumn{1}{l|}{\cellcolor[HTML]{FFFFFF}}                                                    & \multicolumn{1}{l|}{\cellcolor[HTML]{FFFFFF}}                                                    & \cellcolor[HTML]{FFFFFF}                                                    & \multicolumn{1}{l|}{\cellcolor[HTML]{FFFFFF}}                                                    & \multicolumn{1}{l|}{\cellcolor[HTML]{FFFFFF}}                                                     & \multicolumn{1}{l|}{\cellcolor[HTML]{FE0000}}                                                    & \cellcolor[HTML]{FE0000}                                                    \\ \hline
\multicolumn{1}{|l|}{not-co-existence(a,b)}    & \multicolumn{1}{l|}{}                                                                            & \multicolumn{1}{l|}{\cellcolor[HTML]{FFFFFF}}                                                    & \multicolumn{1}{l|}{\cellcolor[HTML]{FE0000}}                                                    & \cellcolor[HTML]{FE0000}{\color[HTML]{FE0000} }                             & \multicolumn{1}{l|}{\cellcolor[HTML]{FE0000}}                                                    & \multicolumn{1}{l|}{\cellcolor[HTML]{FE0000}}                                                     & \multicolumn{1}{l|}{\cellcolor[HTML]{FE0000}}                                                    & \cellcolor[HTML]{FE0000}                                                    \\ \hline
\multicolumn{1}{|l|}{not-succession(a,b)}      & \multicolumn{1}{l|}{}                                                                            & \multicolumn{1}{l|}{\cellcolor[HTML]{FFFFFF}}                                                    & \multicolumn{1}{l|}{\cellcolor[HTML]{FE0000}}                                                    & \cellcolor[HTML]{FE0000}                                                    & \multicolumn{1}{l|}{\cellcolor[HTML]{FE0000}}                                                    & \multicolumn{1}{l|}{\cellcolor[HTML]{FE0000}}                                                     & \multicolumn{1}{l|}{\cellcolor[HTML]{FE0000}}                                                    & \cellcolor[HTML]{FE0000}                                                    \\ \hline
\multicolumn{1}{|l|}{responded-existence(a,b)} & \multicolumn{1}{l|}{}                                                                            & \multicolumn{1}{l|}{\cellcolor[HTML]{FFFFFF}}                                                    & \multicolumn{1}{l|}{\cellcolor[HTML]{FFFFFF}}                                                    & \cellcolor[HTML]{FFFFFF}                                                    & \multicolumn{1}{l|}{\cellcolor[HTML]{FFFFFF}}                                                    & \multicolumn{1}{l|}{\cellcolor[HTML]{FFFFFF}}                                                     & \multicolumn{1}{l|}{\cellcolor[HTML]{FE0000}}                                                    & \cellcolor[HTML]{FFFFFF}                                                    \\ \hline
\end{tabular}
\vspace{-15pt}
\end{table}
  
\subsubsection*{$existence(a)$} $a$ occurs at least once.
\begin{itemize}
    \item {\small $c{=}(\times,\Sigma_1,\Sigma_2) {\nvDash} existence(a)$} if $a {\in} \Sigma_1$ ($a {\in} \Sigma_2$), because for any process tree $M{\in}\mathcal{M}_{c}$, there is a trace $\sigma {\in} \phi(M)$ which only has activities in $\Sigma_2$ ($\Sigma_1$).
    \item {\small $c{=}(\circlearrowleft,\Sigma_1,\Sigma_2) {\nvDash} existence(a)$} if $a {\in} \Sigma_2$, because for any process tree $M{\in}\mathcal{M}_{c}$, there is a trace $\sigma {\in} \phi(M)$ which does not trigger the redo part of the process tree and consists of only activities in $\Sigma_1$.
    \vspace{-10pt}
\end{itemize}

\subsubsection*{$response(a,b)$} If $a$ occurs in the trace, then $b$ occurs as well.
\begin{itemize}
        \item {\small $c{=}(\times,\Sigma_1,\Sigma_2) {\nvDash} response(a,b)$} if $a \in \Sigma_1$ and $b \in \Sigma_2$ (or $a \in \Sigma_2$ and $b \in \Sigma_1$), because for any process tree $M{\in}\mathcal{M}_{c}$, there is a trace $\sigma {\in} \phi(M)$ which has an occurrence of activity $a$ but $b$ cannot occur in this trace.
        \item {\small $c{=}(\wedge,\Sigma_1,\Sigma_2) {\nvDash} response(a,b)$} if $a \in \Sigma_1$ and $b \in \Sigma_2$ (or $a \in \Sigma_2$ and $b \in \Sigma_1$), because for any process tree $M{\in}\mathcal{M}_{c}$, there is a trace $\sigma {\in} \phi(M)$ in which activity $b$ occurs before $a$ and $b$ does not occur again.
        \item {\small $c{=}(\circlearrowleft,\Sigma_1,\Sigma_2) {\nvDash} response(a,b)$} if $a \in \Sigma_1$ and $b \in \Sigma_2$, because for any process tree $M{\in}\mathcal{M}_{c}$, there is a trace $\sigma {\in} \phi(M)$ in which $a$ occurs but $b$ does not occur because the redo part is optional.
        \item {\small $c{=}(\rightarrow,\Sigma_1,\Sigma_2) {\nvDash} response(a,b)$} if $b \in \Sigma_1$ and $a \in \Sigma_2$, because for any process tree $M{\in}\mathcal{M}_{c}$, there is a trace $\sigma {\in} \phi(M)$ in which activity $b$ occurs first and then $a$ occurs and $b$ does not occur again.
         \vspace{-10pt}
    \end{itemize}

    \subsubsection*{$precedence(a,b)$}$b$ occurs only if preceded by $a$.
    \begin{itemize}
        \item {\small $c{=}(\times,\Sigma_1,\Sigma_2) {\nvDash} precedence(a,b)$} if $a \in \Sigma_1$ and $b \in \Sigma_2$ (or $a \in \Sigma_2$ and $b \in \Sigma_1$), because for any process tree $M{\in}\mathcal{M}_{c}$, there is a trace $\sigma {\in} \phi(M)$ which has an occurrence of activity $b$ but $a$ cannot occur in this trace.
        \item {\small $c{=}(\wedge,\Sigma_1,\Sigma_2) {\nvDash} precedence(a,b)$} if $a \in \Sigma_1$ and $b \in \Sigma_2$ (or $a \in \Sigma_2$ and $b \in \Sigma_1$), because for any process tree $M{\in}\mathcal{M}_{c}$, there is a trace $\sigma {\in} \phi(M)$ in which activity $b$ occurs before $a$.
        \item {\small $c{=}(\circlearrowleft,\Sigma_1,\Sigma_2) {\nvDash} precedence(a,b)$} if $b \in \Sigma_1$ and $a \in \Sigma_2$, because for any process tree $M{\in}\mathcal{M}_{c}$, there is a trace $\sigma {\in} \phi(M)$ in which $b$ occurs but $a$ does not occur because the redo part is optional.
        \item {\small $c{=}(\rightarrow,\Sigma_1,\Sigma_2) {\nvDash} precedence(a,b)$} if $b \in \Sigma_1$ and $a \in \Sigma_2$, because for any process tree $M{\in}\mathcal{M}_{c}$, there is a trace $\sigma {\in} \phi(M)$ in which activity $b$ occurs first and then $a$ occurs.
        \vspace{-10pt} 
    \end{itemize}

\subsubsection*{$co \mhyphen existence(a,b)$} $a$ and $b$ occur together.
\begin{itemize}
    \item {\small $c{=}(\times,\Sigma_1,\Sigma_2) {\nvDash} co \mhyphen existence(a,b)$ if $a \in \Sigma_1$} and $b \in \Sigma_2$ (or $a \in \Sigma_2$ and $b \in \Sigma_1$), because for any process tree $M{\in}\mathcal{M}_{c}$, there is a trace $\sigma {\in} \phi(M)$ which has an occurrence of only activity $a$ or only activity $b$.
    \item {\small $c{=}(\circlearrowleft,\Sigma_1,\Sigma_2) {\nvDash} co \mhyphen existence(a,b)$ if $a \in \Sigma_1$} and $b \in \Sigma_2$ ($a \in \Sigma_2$ and $b \in \Sigma_1$), because for any process tree $M{\in}\mathcal{M}_{c}$, there is a trace $\sigma {\in} \phi(M)$ in which only activity $a$ ($b$) occurs because the redo part is optional.
     \vspace{-10pt}
    \end{itemize}

\subsubsection*{$not \mhyphen co \mhyphen existence(a,b)$} $a$ and $b$ cannot occur together.
\begin{itemize}
        \item {\small $c{=}(\wedge,\Sigma_1,\Sigma_2) {\nvDash} not \mhyphen co \mhyphen existence(a,b)$} if $a \in \Sigma_1$ and $b \in \Sigma_2$ (or $a \in \Sigma_2$ and $b \in \Sigma_1$), because for any process tree $M{\in}\mathcal{M}_{c}$, there is a trace $\sigma {\in} \phi(M)$ in which both activities $a$ and $b$ occur.
        \item {\small $c{=}(\circlearrowleft,\Sigma_1,\Sigma_2) {\nvDash} not \mhyphen co \mhyphen existence(a,b)$} if $a \in \Sigma_1$ and $b \in \Sigma_2$ (or $a \in \Sigma_2$ and $b \in \Sigma_1$), because for any process tree $M{\in}\mathcal{M}_{c}$, there is a trace $\sigma {\in} \phi(M)$ in which $a$ and $b$ both occur because the redo part is optional and may occur. Also, if $a \in \Sigma_1$ and $b \in \Sigma_1$ (or $a \in \Sigma_2$ and $b \in \Sigma_2$), since the do part or redo part of the loop tree can occur multiple times and the repetitions are independent, then activities $a$ and $b$ may occur in different repeats of the loop process trees.
        \item {\small $c{=}(\rightarrow,\Sigma_1,\Sigma_2) {\nvDash} not \mhyphen co \mhyphen existence(a,b)$} if $a \in \Sigma_1$ and $b \in \Sigma_2$ (or $a \in \Sigma_2$ and $b \in \Sigma_1$), because for any process tree $M{\in}\mathcal{M}_{c}$, there is a trace $\sigma {\in} \phi(M)$ in which both activities $a$ and $b$ occur.
        \vspace{-10pt} 
    \end{itemize}

\subsubsection*{$not\mhyphen succession(a,b)$} $b$ cannot occur after $a$.
\begin{itemize}
        \item {\small $c{=}(\wedge,\Sigma_1,\Sigma_2) {\nvDash} not\mhyphen succession(a,b)$} if $a \in \Sigma_1$ and $b \in \Sigma_2$ (or $a \in \Sigma_2$ and $b \in \Sigma_1$), because for any process tree $M{\in}\mathcal{M}_{c}$, there is a trace $\sigma {\in} \phi(M)$ in which activity $a$ occurs first and then $b$ occurs after it.
        
         \item {\small $c{=}(\circlearrowleft,\Sigma_1,\Sigma_2) {\nvDash} not\mhyphen succession(a,b)$} if $\{a,b\} \subseteq \Sigma_1 \cup \Sigma_2$, because for any process tree $M{\in}\mathcal{M}_{c}$, there is a trace $\sigma {\in} \phi(M)$ in which activity $b$ occurs after activity $a$. 
         Since the do part and redo part of the loop tree can occur multiple times and the repetitions are independent, then after the occurrence of activity $a$, activity $b$ can eventually occur.

        \item {\small $c{=}(\rightarrow,\Sigma_1,\Sigma_2) {\nvDash} not\mhyphen succession(a,b)$} if $a \in \Sigma_1$ and $b \in \Sigma_2$, because for any process tree $M{\in}\mathcal{M}_{c}$, there is a trace $\sigma {\in} \phi(M)$ in which activity $a$ occurs first and then $b$ occurs.
        \vspace{-10pt} 
    \end{itemize}
 
\subsubsection*{$responded\mhyphen existence(a,b)$} If $a$ occurs in the trace, then $b$ occurs as well.
\begin{itemize}
    \item {\small $c{=}(\times,\Sigma_1,\Sigma_2) {\nvDash} responded\mhyphen existence(a,b)$} if $a \in \Sigma_1$ and $b \in \Sigma_2$ (or $a \in \Sigma_2$ and $b \in \Sigma_1$), because for any process tree $M{\in}\mathcal{M}_{c}$, there is a trace $\sigma {\in} \phi(M)$ which has an occurrence of only activity $a$ and $b$ cannot occur in this trace.
    \item {\small $c{=}(\circlearrowleft,\Sigma_1,\Sigma_2) {\nvDash} responded\mhyphen existence(a,b)$} if $a \in \Sigma_1$ and $b \in \Sigma_2$, because for any process tree $M{\in}\mathcal{M}_{c}$, there is a trace $\sigma {\in} \phi(M)$ in which only activity $a$ occurs but $b$ does not occur since the redo part is optional.
    \end{itemize}

\vspace{-20pt}
\section{Evaluation}
The proposed framework is implemented and is publicly available\footnote{\url{https://github.com/aliNorouzifar/IMr}}. We used several real-life event logs to evaluate the framework including BPIC 2012 (application and offer sub-processes), BPIC 2017 (application and offer sub-processes), BPIC 2018 (application sub-process), Hospital Billing, Sepsis, and UWV event logs. All the event logs except UWV are publicly available\footnote{\url{https://data.4tu.nl/}}. Key statistics for these event logs are summarized in Table~\ref{event_logs}. The evaluation section is structured into two parts. Initially, we present results derived from publicly available event logs. Subsequently, we offer insights from a real-life case study conducted in collaboration with UWV, the Dutch employee insurance agency. Domain experts at UWV actively contributed to extracting a normative model for a claim-handling process and validating the obtained results.

\begin{table}[tb]
\scriptsize
\caption{Event logs used in experiments.}
\label{event_logs}
\centering
\begin{tabular}{c|c|c|c|c|}
\cline{2-5}
                               & \#activities & \#events  & \#traces & \#trace variants \\ \hline
\multicolumn{1}{|c|}{BPIC 12}  & 17           & 92,093    & 13,087   & 576              \\ \hline
\multicolumn{1}{|c|}{BPIC 17}  & 18           & 433,444   & 31,509   & 2,630            \\ \hline
\multicolumn{1}{|c|}{BPIC 18}  & 15           & 928,091   & 43,809   & 1,435            \\ \hline
\multicolumn{1}{|c|}{Hospital} & 18           & 451,359   & 100,000  & 1,020            \\ \hline
\multicolumn{1}{|c|}{Sepsis}   & 16           & 15,214    & 1,050    & 846              \\ \hline
\multicolumn{1}{|c|}{UWV}      & 16           & 1,309,719 & 144,046  & 484              \\ \hline
\end{tabular}
\vspace{-15pt}
\end{table}

\vspace{-10pt}
\subsection{Real-life Event Logs}
Our framework is designed independently from the source which provides the rules. We use Declare Miner \cite{DBLP:conf/caise/MaggiBA12} with the subset of declarative templates introduced in this paper and select the rules with $confidence = 1$, i.e., the constraints for which there is no trace in the event log that deviates. However, one could employ various heuristics to account for noisy behavior or other considerations. Our initial comparison involves assessing models discovered by IMf \cite{DBLP:conf/bpm/LeemansFA13}, a state-of-the-art algorithm, against those produced by the IMr algorithm utilizing discovered declarative constraints. Experiments are conducted with the infrequency parameter $f$ ranging from 0 to 1 in intervals of 0.1. In the IMr algorithm, we vary the $sup$ parameter within the range of 0 to 1 at intervals of 0.1.

\begin{figure}[tb]
\centering
\begin{subfigure}{0.49\textwidth}
\includegraphics[scale=0.35]{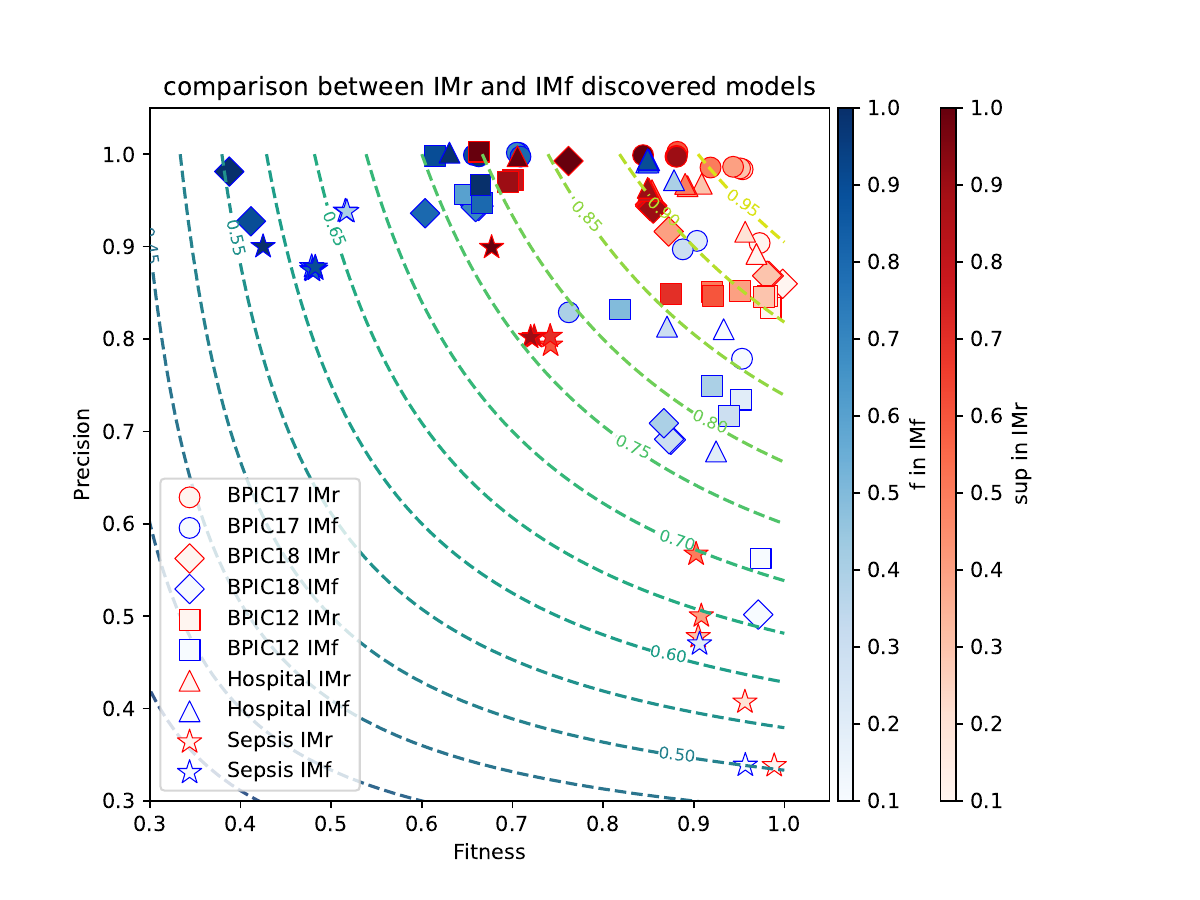}
\caption{\small Comparison between process models discovered using IMr and IMf.}
\label{compare_IMr_IMf}
\end{subfigure}\;
\begin{subfigure}{0.49\textwidth}
\centering
\includegraphics[scale=0.35]{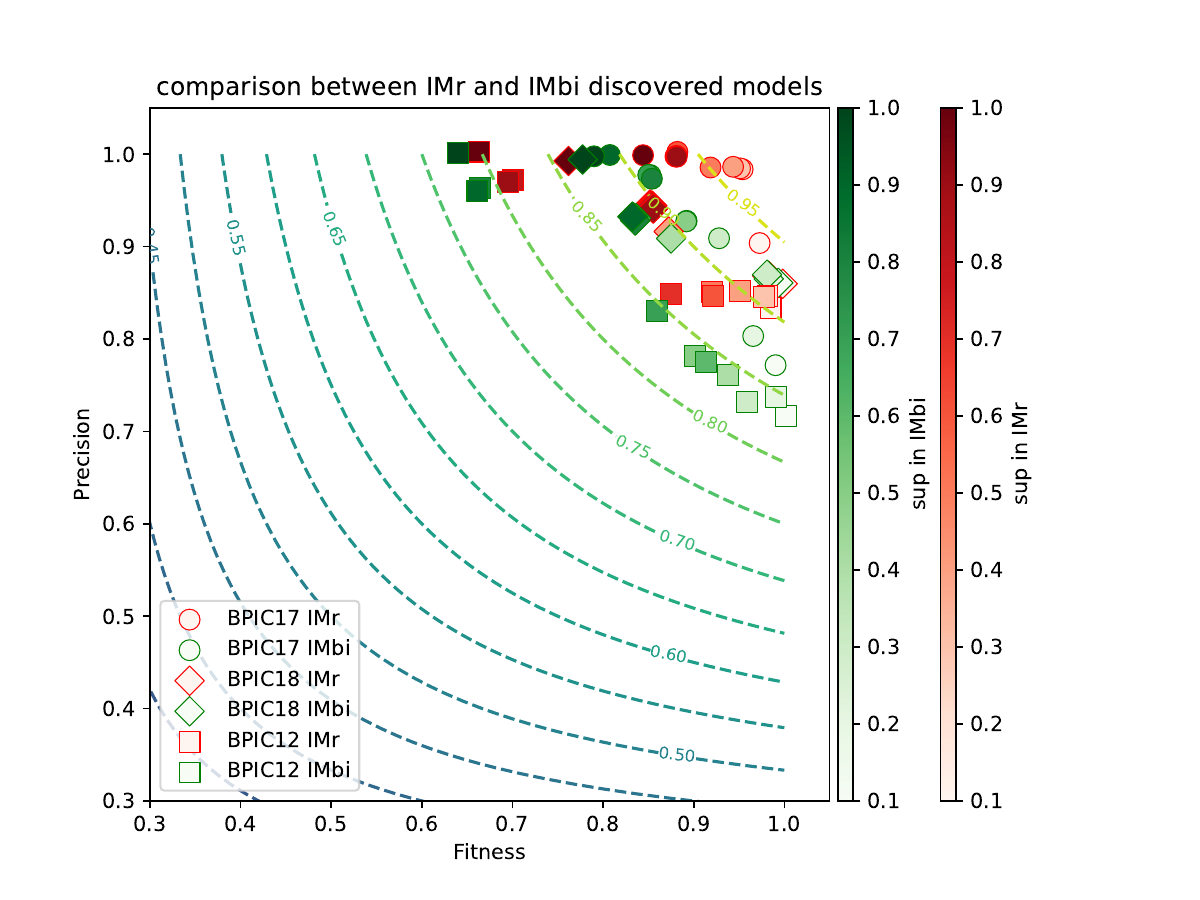}
\caption{\small Comparison between process models discovered using IMr and IMbi.}
\label{compare_IMr_IMbi}
\end{subfigure}
% \label{compare_IMr_IMf}
\caption{\small Comparison between process models discovered using IMr, IMf, and IMbi.As can be seen, IMr (red shapes) models perform better than IMbi (green shapes) and IMf (blue shapes) models.}
\vspace{-20pt}
\end{figure}

The visual representation of the discovered models' quality is depicted in Fig.~\ref{compare_IMr_IMf}, employing well-known evaluation metrics. Specifically, the x-axis represents alignment fitness \cite{CarmonaDSW18}, the y-axis represents precision \cite{CarmonaDSW18}, and the contours on the plot illustrate the F1-score derived from these two values. Various shapes differentiate between event logs, i.e., $\bigcirc$: BPIC17, $\Diamond$: BPIC18, $\square$: BPIC12, $\triangle$: Hospital, and $\largestar$: Sepsis, shapes with blue color represent IMf models and shapes with red color represent IMr models while color intensity corresponds to the parameters of the discovery algorithms, i.e., $f$ in IMf and $sup$ in IMr. Notably, the figures indicate that, in general, IMr models outperform IMf models across the three evaluation metrics illustrated in the plot.

In Fig.~\ref{compare_IMr_IMbi}, the models discovered with IMr are shown with red color and the models discovered with IMbi are shown with green color. For the Hospital Billing and Sepsis event logs, it was infeasible to discover process models using the IMbi algorithm considering a maximum run time of one hour. The discovered model using BPIC 12 and BPIC 17 shows that IMr models score better. For BPIC 2018, although the discovered models are very similar, the run time of the IMr algorithm is about four times shorter.

Experiments exceeding the one-hour maximum run-time were terminated. While the IMf algorithm is notably fast, delivering a model within seconds, the computational costs for IMbi and IMr are considerably higher. In Table~\ref{run_time}, some statistics are presented to compare the run time of IMbi with IMr\footnote{The time required for rule extraction is not included.}. Notably, a comparison between the number of candidate cuts in the initial recursion of both IMbi and IMr demonstrates the effect of utilizing rules in efficiently pruning the search space while preserving model quality, as indicated in Fig.~\ref{compare_IMr_IMbi}. The candidate cut search is independent of the $sup$ parameter. The table additionally includes information on the average run-time duration for each event log that shows IMr is considerably faster because of the reduced number of candidate cuts.

\begin{table}[tb]
\scriptsize
\caption{\small Run time statistics IMr and IMbi}
\label{run_time}
\centering
\begin{tabular}{|l|c|c|c|c|c|}
\hline
                      & BPIC 12 & BPIC 17 & BPIC 18 & Hospital   & Sepsis    \\ \hline
$\vert C \vert$ in the first iteration of IMbi  & 3601    & 4659    & 8103    & 106771     & 153502    \\ \hline
$\vert C \vert$ in the first iteration of IMr   & 12      & 47      & 19      & 8          & 329       \\ \hline
average run time of IMbi & 20 sec.      & 176 sec.     & 801 sec.     & $> 1$ hour & $>1$ hour \\ \hline
average run time of IMr  & 11 sec.      & 55 sec.      & 201 sec.     & 152 sec.       & 9 sec.         \\ \hline
\end{tabular}
\end{table}

\begin{figure}[tb]
\centering
\includegraphics[scale=0.35]{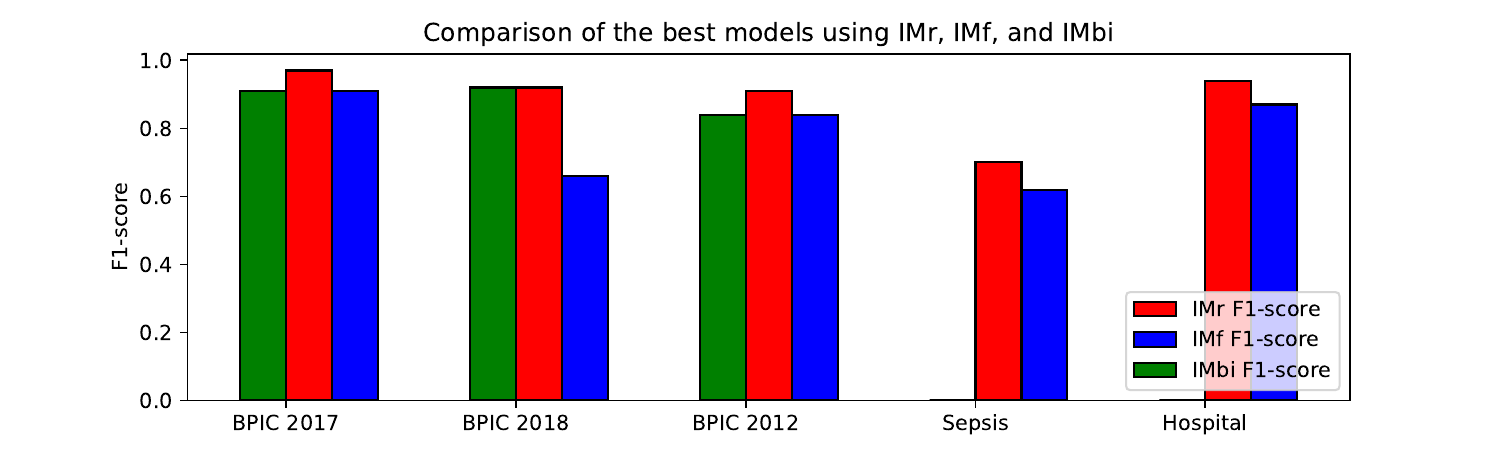}
\caption{\small Comparison between best models discovered by IMr, IMf, and IMbi (for Sepsis and Hospital event logs it was not feasible to discover a model in an hour using IMbi).}
\label{bestmodels_IMr_IMbi}
\vspace{-20pt}
\end{figure}

In Fig.~\ref{bestmodels_IMr_IMbi}, we chose the best models for each event log using different parameters in IMf, IMr, and IMbi experiments. The criterion for selecting the best model involves choosing the one with the highest F1-score among those with a minimum alignment fitness of 0.9. The bar chart in this figure compares alignment fitness, precision, and F1-score for the selected models.

\begin{figure}[htb]
\begin{subfigure}{1\textwidth}
\centering
\includegraphics[scale=0.15]{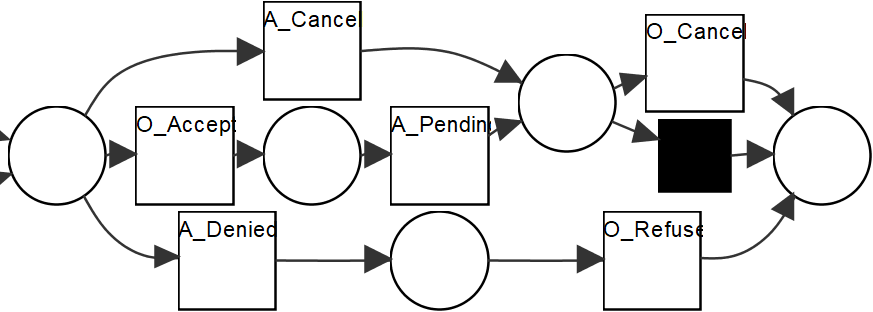}
\caption{\small A part of the discovered model for BPIC 2017 event log using IMr with $sup=0.2$.}
\label{motive_IMr_2017}
\end{subfigure}\\
\begin{subfigure}{1\textwidth}
\centering
\includegraphics[scale=0.15]{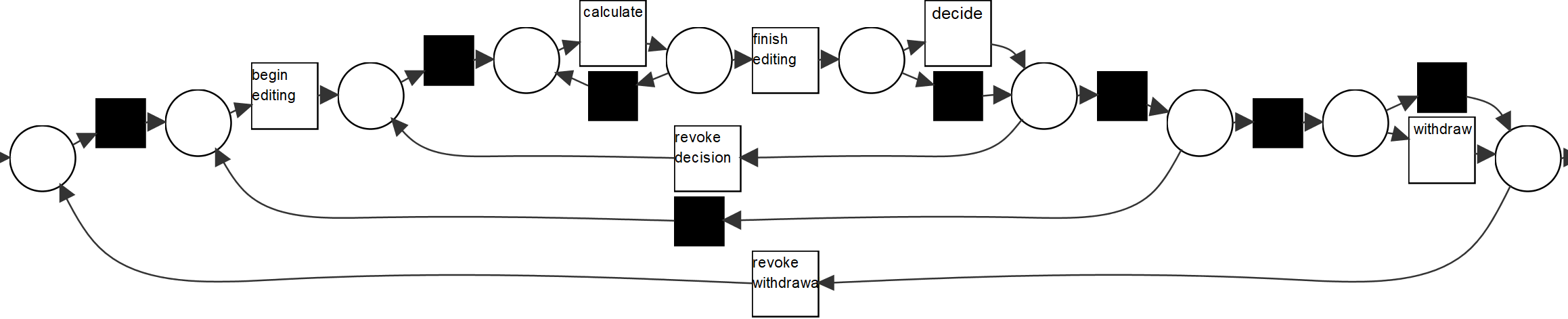}
\caption{\small A part of the discovered model for BPIC 2018 event log using IMr with $sup=0.1$.}
\label{motive_IMr_2018}
\end{subfigure}
\caption{Discovered models using IMr for BPIC 2017 and BPIC 2018}
\label{motiv_example_after}
\vspace{-10pt}
\end{figure}

In Fig.\ref{motiv_example_before}, we provided examples to motivate our goal. To emphasize the impact of our proposed IMr framework, a part of the discovered models is presented in Fig.\ref{motiv_example_after}. In Fig.\ref{motive_IMr_2017}, it is evident that only one of the activities \textit{A\_Cancelled}, \textit{A\_Denied}, or \textit{A\_Pending} can occur. In Fig.\ref{motive_IMr_2018}, the sequential order of transitions \textit{begin editing}, \textit{calculate}, \textit{finish editing}, and \textit{decide} is more coherent, aligning seamlessly with our understanding of the process.

\vspace{-10pt}

\vspace{-20pt}
\subsection{Case Study UWV}
\begin{figure}[htb]
\centering
\includegraphics[scale=0.1]{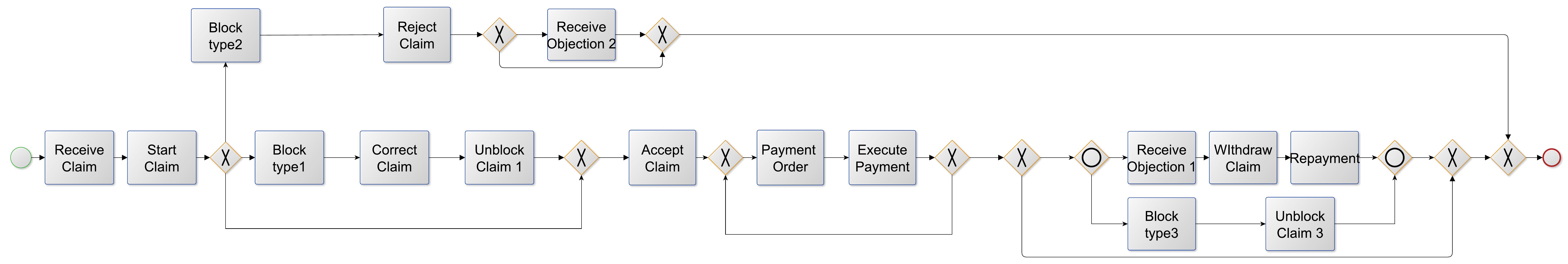}
\caption{\small UWV normative model.}
\label{normative_model_uwv}
\vspace{-10pt}
\end{figure}

This case study is a collaborative effort with the UWV agency, involving the analysis of an event log pertaining to one of their claim-handling processes, encompassing data for 144,046 clients. Fig.~\ref{normative_model_uwv} illustrates the normative model derived from a comprehensive examination of the event log, supplemented by insights from process experts who actively contributed their domain knowledge throughout this case study. After a claim is received the process is started, some cases are blocked (\textit{Block type 1}) and consequently, some corrections are performed and the case is unblocked afterward. Some other cases after starting a claim have another type of blocking (\textit{Block type 2}) which is followed by rejecting the claim and possibly an objection from the client. If the case is accepted, then the client is entitled to receive some payments (between one to three payments). Some clients file an objection after receiving the payments which continues with withdrawing the claim and repaying the received money to UWV. Optionally a third type of blocking might also occur (\textit{Block type 3}) to prevent any pending payments to the client from being made.

IMf and IMr with different parameter settings are utilized to discover process models from the UWV event log. The set of rules $R$ extracted from the event log using Declare Miner \cite{DBLP:conf/caise/MaggiBA12} with $confidence{=}1$ is used in IMr. Some examples of discovered declarative constraints are {\small $precedence(Block\; type\; 1,Unblock\; Claim\; 1)$}, {\small $response(Block\; type\; 2, Reject\; Claim)$}, {\small $not\mhyphen succession(Execute\; Payment,Accept\; Pa\mhyphen$ $yment)$}, and {\small $not \mhyphen co \mhyphen existence(Block\; type\; 1,$ $ Block\; type\; 2)$}. These rules align with the normative model and can be used to guide IMr to discover better models.

\begin{figure}[htb]
\centering
\includegraphics[scale=0.35]{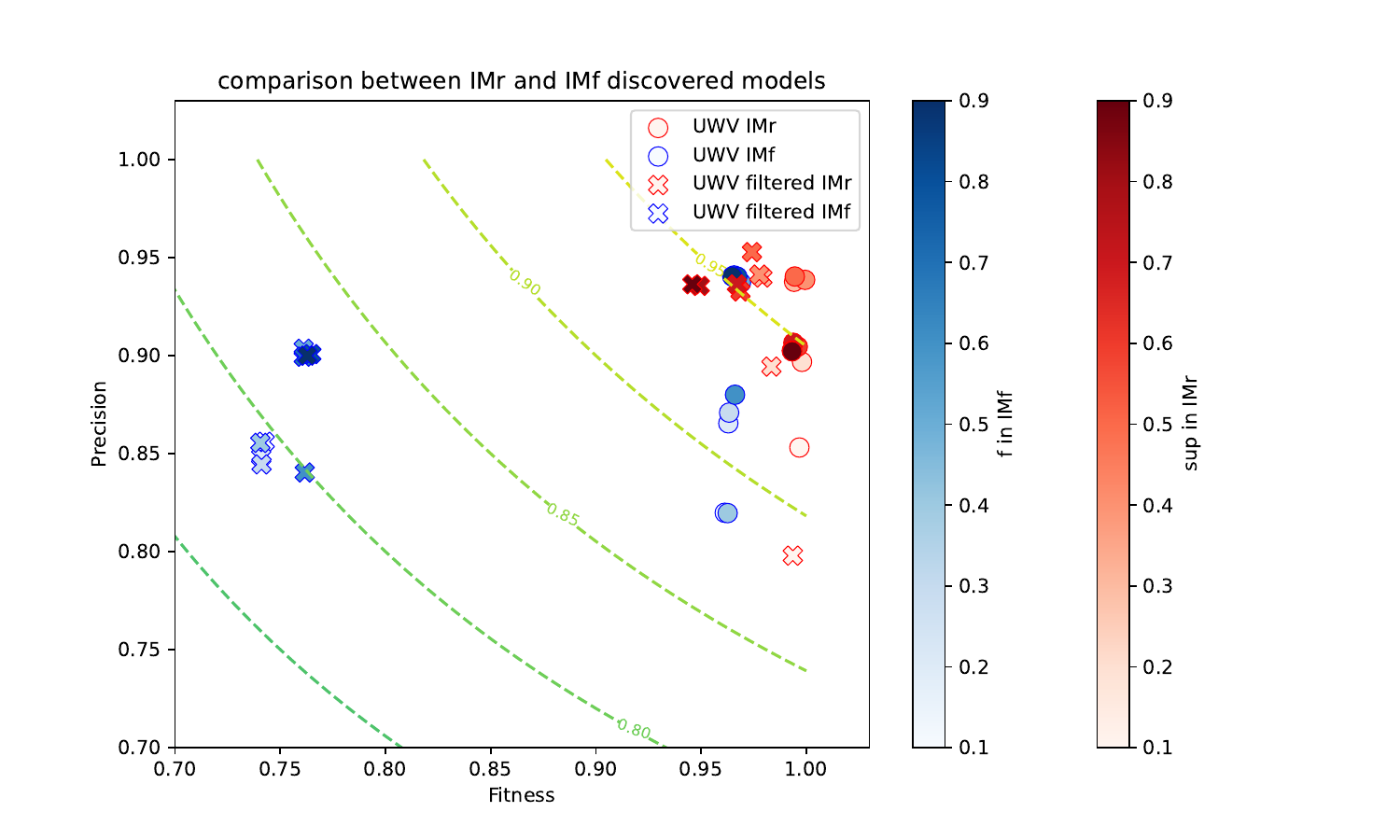}
\caption{\small Comparison between process models discovered for UWV event log using IMr and IMf.}
\label{compare_IMr_IMf_uvw}
\vspace{-20pt}
\end{figure}

A comparative analysis of these models is presented in Fig.~\ref{compare_IMr_IMf_uvw}. The circle shape $\bigcirc$ in the figure represents the original event log without any filtering applied. Although IMr models exhibit superior scores, the difference does not seem significant. This is primarily attributed to the most frequent trace variant, constituting 86\% of the data, significantly influencing the alignment fitness value. Both IMr and IMf discovered models replay this trace variant. However, upon filtering out this prevalent trace variant, the contrast becomes clearer, evident in the figure with cross-shaped points $\mathbb{X}$. The IMf models have lower fitness values which shows their difficulties in modeling these trace variants. In contrast, IMr models show a robust representation of the process, as the removal of this frequent trace variant has a marginal impact on the overall model quality.

\begin{figure}[htb]
\centering
\includegraphics[scale=0.13]{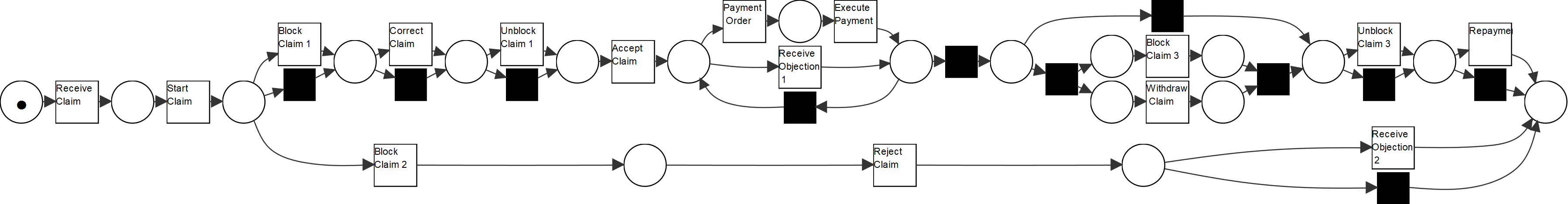}
\caption{\small Discovered model for UWV event log using IMr with $sup=0.5$.}
\label{uwv_model_after}
\vspace{-20pt}
\end{figure}

It was infeasible to discover process models in one hour using IMbi, therefore the comparison between IMbi models and IMr models is excluded from the experiments. The average run-time of the IMr algorithm in different experiments is 348 seconds. The best model discovered from the complete event log with IMf using $f=0.5$ has an alignment fitness of 96.7, a precision of 93.9, and an F1-score of 95.3. The best model discovered with IMr using $sup=0.5$ has an alignment fitness of 99.6, a precision of 93.8, and an F1-score of 96.6. This is illustrated in Fig.~\ref{uwv_model_after}. This model represents the process much better than Fig.~\ref{IMf_model_uwv}, especially if we compare it with the normative model illustrated in Fig.~\ref{normative_model_uwv}.

\section{Open Challenges}
\label{open_chal}
Due to the representational bias of inductive mining algorithms, the IMr algorithm does not take into account long-term dependencies in the discovery procedure. However, declarative rules may represent long-term dependencies. Consider the event log $L_1=[\langle a,c,d \rangle, \langle b,c,e \rangle]$ and set of rules 
{\small $R_1{=}\{not \mhyphen co \mhyphen existence(a,e),$ $not \mhyphen co \mhyphen existence(b,d)\}$}. All traces in $L_1$ satisfy all the rules in $R_1$. In the first recursion of the IMbi algorithm, the set of possible cuts without considering the rules in $R_1$ is {\small
$C=\{
\rightarrow(\{a \}, \{b,c,d,e \}),
\rightarrow(\{b \}, \{a,c,d,e \}),
\rightarrow(\{a,b \}, \{c,d,e \}),
\rightarrow(\{a,b,c \}, \{d,e \}),
\rightarrow(\{a,b,c$ $,d \}, \{e \}),
\rightarrow(\{a,b,c,e \}, \{d \})\}$}.
If we use IMr with the set of rules $R_1$, because of the long-term dependencies between $a$ and $e$, and between $b$ and $d$, all the cuts are rejected. Applying any of these cuts results in traces that do not satisfy at least one rule. In such cases, IMr continues with ignoring the set of rules in that specific recursion.

Consider the event log $L_2=[\langle c,a,c,b,c \rangle]$, and set of rules {\small $R_2{=}\{response(a,b)\}$}.  The dependency between $a$ and $b$ is long-term and observed in different runs of a loop. Considering $sup=0.2$, the first recursion of IMr finds $\circlearrowleft(\{c \}, \{a,b \})$ which splits the event log as $L_3=[\langle c \rangle^3]$ and $L_4=[\langle a \rangle, \langle b \rangle]$. The only possible candidate cut in $IMr(L_4,0.2,R_2)$ is $\times(\{a\}, \{b\})$ which is rejected based on $R_2$. The dependency is between different runs of the process. The discovered model using IMbi without considering the rules is $\circlearrowleft(\{c \}, \times(\{a\}, \{b\}))$ that can generate traces $\langle c,a,c \rangle$, and $\langle c,b,c,a,c \rangle$ which deviates $response(a,b)$.

 Our framework always guarantees a sound model. The declarative rules in IMr are used to guide the algorithm to understand the order of activities. Therefore, it cannot guarantee that the final model satisfies all the input rules. For example, consider the event log $L_5{=}[\langle a,c,b \rangle^{50}, \langle d,c \rangle^{50}]$ and $R_3{=}\{precedence(a,b) \}$. IMr with $sup=0.2$ and $R_3$ discovers $\rightarrow (\times(a,d),\rightarrow(c,\times(b,\tau)))$. This model allows for trace $\langle d,c,b \rangle$ which violates $precedence(a,b)$. In case the strict version of the algorithm is used which outputs no model if all cuts are rejected in a recursion, we provide the following guarantees:   $at \mhyphen most(a)$: in all traces, $a$ can occur at most one time,
     $existence(a)$: in all traces, $a$ can occur,
     $response(a,b)$: each time $a$ occurs, $b$ can occur after it,
     $precedence(a,b)$: each time $b$ occurs, it is possible that $a$ occurred before it,
     $co\mhyphen existence(a,b)$: $a$ and $b$ can occur together,
     $not\mhyphen co\mhyphen existence(a,b)$: $a$ and $b$ never occur together,
     $not\mhyphen succession(a,b)$: $b$ cannot occur after $a$, 
     $responded\mhyphen existence(a,b)$: if $a$ occurs in the trace, then $b$ can occur as well.

\section{Conclusion}
The proposed framework is based on the inductive mining methodology and makes a contribution to the field by introducing a novel variant of inductive mining. This variant incorporates the use of user-defined or discovered rules during the process discovery recursions. Through extensive evaluation, our results demonstrate that the discovered models surpass current approaches, yielding more accurate process models that align closely with process knowledge available prior to process discovery. This framework holds the potential for extension into an interactive process discovery framework. In such a setting, domain experts can interactively examine discovered models, utilizing a set of rules to guide the discovery algorithm toward refining the process model. This iterative process enhances the adaptability of the framework and ensures alignment with evolving domain expertise. Moreover, the discussed approach offers the possibility of handling scenarios involving multiple event logs. The framework can be extended to discover a process model that supports a desirable event log while simultaneously avoiding an undesirable event log. 

\bibliographystyle{splncs04}
\bibliography{lit}

\end{document}